\documentclass[submission,creativecommons]{eptcs}
\providecommand{\event}{AISOS 2013} 
\usepackage{breakurl}             

\title{Challenges for modelling and analysis in embedded systems and systems-of-systems design}
\author{Boudewijn R. Haverkort
\institute{Centre for Telematics and Information Technology\\
University of Twente\thanks{This paper has been inspired by the experience gained while I was scientific director of the Embedded Systems Institute (ESI), in the years 2009--2012. This paper expresses my personal opinion, and not necessarily that of ESI, nor of the companies ESI is or has been cooperating with.}\\
P.O.~Box 217, 7400 AE Enschede, the Netherlands}
\email{b.r.h.m.haverkort@utwente.nl}
}
\def\titlerunning{Challenges for modelling and analysis in embedded systems and systems-of-systems design}
\def\authorrunning{Boudewijn R. Haverkort}
\begin{document}
\maketitle

\begin{abstract}
Over the last decade we have witnessed an increasing use of data processing in embedded systems. Where in the past
the data processing was limited (if present at all) to the handling of
a small number of ``on-off control signals'', more recently much more
complex sensory data is being captured, processed and used to
improve system performance and dependability. The advent of
systems-of-systems aggravates the use of more and more data, for
instance, by bringing together data from several independent sources, allowing, in
principle, for even better performing systems. However, this ever
stronger data-orientation brings along several challenges in system
design, both technically and organisationally, and also forces
manufacturers to think beyond their traditional field of
expertise. In this short paper, I will address these new design challenges, through a number of examples. The paper finishes with concrete challenges for supporting tools and techniques for system design in this new context.
\end{abstract}

\section{Introduction}

Over the last decade we have witnessed an increasing use of data
measurements and data processing in embedded systems. Where in the past
the data processing was limited, if present at all, to the handling of
a small number of ``on-off control signals'', more recently much more
complex sensory data is being captured, processed and used to
improve system performance and dependability. In a recent roadmap discussion, it was stated that  ``a 10 km car drive these days, produces and uses more data than NASA used to put Armstrong on the moon''.  Of course, this has been made possible by the dramatic increase of capacity and reduction of price of processing and memory elements (driven by Moore's law), in combination with a similar evolution in sensor equipment. Why do car manufacturers put so much ICT in their cars? Well, to make them perform better, make them more reliable, safer, more efficient, more comfortable, and less polluting. Without embedded ICT in cars, it would be (almost) impossible to fulfil EU exhaust-regulations.

Modern cars are an example of complex high-tech systems in which the embedded ICT component plays a key role. Other sectors, next to automotive, where this is similar are, for instance, healthcare, energy, professional printing, manufacturing, distribution and logistics, situational awareness, avionics, and defence. In all these sectors, OEM's (original equipment manufacturers) are facing dramatic changes in the way their products are being developed: where in the past ``steel, oil and rubber'' were the main ingredients, more and more, ICT is determining the functionality, performance and competitiveness of their products. Although manufacturers do typically not openly publish these numbers, claims that 40-50\% of the development costs of a new car are related to ICT are not uncommon. Almost invisibly, over the years, many traditional high-tech companies have become ICT companies; the only difference with classical ICT companies, like Microsoft, Oracle or SAP,  is that the user interfacing is different. The uprise of so-called systems-of-systems does make the role of ICT even more important. On top of that, other challenges in systems-of-systems (see below), make the design of correctly functioning systems-of-systems even more a challenge. 

In what follows, I well briefly touch upon different system classes in Section~\ref{system-types}, followed by a number of examples in Section~\ref{examples}, illustrating the increasing role of data. In Section~\ref{challenges}, I will then discuss a number of challenges that follow from this regarding design of these systems. These, in turn, set challenges for the tools and techniques that are needed to support these design processes.

\section{System types}
\label{system-types}

One could argue that the examples from the automotive domain given above are {\em embedded systems}, rather than {\em systems-of-systems}. An embedded system is typically seen as ``a computer system (hardware and software) designed to interact with the physical world; it is embedded as part of a complete system, including sensors and actuators''. With the term embedded systems, most people think of ``systems you buy in a box''. The examples given above are primarily of that type. However, infrastructural systems, like systems for surveillance or traffic control,  even though they are much more geographically spread, share a lot of the characteristics with embedded systems. With such systems, however, we are entering the realm of systems-of-systems. Before coming to these, however, is is important to briefly address the notion of {\em cyber-physical systems}, as coined in various NSF workshops \cite{NFS-CPS,stankovic2005}. Where in these workshops the aspects of control and communications were very much stressed as distinguishing features (as opposed to embedded systems), I do think that this new term is largely a matter of taste. In the EU Artemis program (see \url{http://www.artemis-ia.eu/})\footnote{All URL's in this paper have been validated on July 17, 2013.} the term normally employed is embedded systems, however, the systems being addressed do heavily use communication and do employ (or require) a large variety of control, hence, Artemis is addressing cyber-physical systems as well.

But what do systems-of-systems then really add? What does make them different? At this point, it is instructive to go back to the original description put forward by Maier, already 15 years ago \cite{maier1998}: {\em a system-of-systems is an assemblage of components which may individually be regarded as systems themselves, with two additional properties: }
\begin{itemize}
\item {\em operational independence: disassembled components must be able to do useful work independently, and}
\item {\em managerial independence: disassembled  components do work independently.}
\end{itemize}
Importantly, component systems can have different ownership and can underlie different legislation. 

The question you might ask then: is this more complex than ordinary embedded systems? The answer is a clear yes, for various reasons. Indeed, a system-of-system is an assemblage of systems, integrated out of independent components {\em as they are}, take it or leave it. That is, when integrating the subsystems, in no reasonable way, adaptations to the component-systems can be made. Furthermore, it might require run-time adaptation of components, as the different components comprising the overall system might be updated or changed (within reasonable bounds, of course), thus requiring adaptations form the surrounding subsystems. Here, in essence, an online integration and test capability, normally only part of the off-line design process, is required. And knowing how difficult and time-consuming normal integration and test already is, this clearly puts an extra challenge. Next to that, intuitively, the black-box character of the components being assembled is very high. This makes putting them together in such a way that guarantees can be given with respect to extra-functional properties like dependability or performance is extremely difficult. Not to mention security issues. Ways of working involving service-level agreements (SLA's) like done in some networking or cloud computing solutions appear to be appropriate here. But do note that many systems-of-systems, unlike most internet applications, are employed for applications with real-time characteristics, making this even more challenging.

\section{Examples}
\label{examples}

\subsection{The internet}
Probably the best known system-of-systems is the internet, in which many independent internet service providers (ISPs) are cooperatively providing a world-wide network coverage, on the basis of jointly agreed interfaces and protocols. The internet as we know it today has been developed since the beginning of the 1980's, without any notion of systems-of-systems being around. {\em Within} the domain of one ISP, that ISP has freedom to choose its own implementation to a certain extend, for instance for routing, as long as it adheres to the externally agreed-upon service levels and interfaces. Note, however, that the internet, at that level (network layer) is a best effort network, that is, a system that does not fulfil real-time requirements. Lessons can be learned from the internet context, but surely, more is needed for systems-of-systems. For more background on the internet, refer to \cite{KR2012}.

\subsection{Cooperative adaptive cruise control}
We now consider a cooperative adaptive cruise control (CACC) system, as, for instance, worked upon in the Dutch Connect \& Drive project \cite{Ploeg2011} (see also \url{http://www.youtube.com/watch?v=OoRuE7OqFEs}). In a car equipped with a CACC system, the car its cruise control is not only fed with the usual controls from within the car (set-point and some car-internal sensor readings), but also reacts and adapts on signals from the outside world, e.g., by making use of radar or infrared communication to measure the distance to preceding cars. What we did specifically in the mentioned project, however, was also exchanging information using WLAN (specifically, IEEE 802.11p) between cars in near proximity. In doing so, cars can easily exchange information (digitally) about their position (on the basis of GPS readings), speed and acceleration. By forwarding information of cars ``in front'' to cars ``behind'', in essence cars look further ahead  than is possible on the basis of just radar of infrared communication. Next to that, cars could also communicate with road-side stations, or even subscribe to traffic services providing detailed information about traffic situations ahead. In any case, the information received should be trustworthy, and also not be outdated, otherwise it becomes a safety risk to base speed adaptation on it. Next to that, not under all circumstances will all this information channels be active (due to failures or other reasons beyond control of any single car). Hence, robustness, that is, the ability to deal with unplanned or undesirable future circumstances, is a key requirement.

Note that the cars are each privately owned and will be of different brands. GPS is provided publicly, as are road-side electronic information systems. Traffic services are typically privately (commercially) provided, using mobile telephony channels (3G or UMTS) by one or multiple providers. The overall system provides, as new emerging service, a much better, smoother traffic flow. The CACC system as a whole is made up of differently owned components, and different legislation might be involved. The question does arise, actually, who is at the steering wheel? In case of accidents due to the receipt of incorrect information through one of the data channels, liability questions will arise!

\subsection{Integrated fleet control and optimisation}
Staying in the automotive area, one can also think of fully integrated fleet management systems, in at least two ways, as follows.

First of all, in modern cars, embedded systems are present for engine and exhaust control. In hybrid cars this can be combined with real-time route planner information systems, to plan electrical engine usage such that full advantage is taken from the traffic conditions and the changing heights levels in the route. To save as much energy as possible, the engine control system strives to use as much electrical energy when driving up hill, so that the battery is (almost) empty when arriving at the top, so that a maximum amount of energy can be regained when going down using the electrical drive train as energy-generating brake. 

Secondly, when all cars in a fleet report digitally, e.g., via 3G public telephony, their routing and all kinds of other sensory information to a central location, e.g., the headquarters of a transport company, real-time information is available about the whereabout and status of the complete fleet, thus making adaptive trip planning possible. Furthermore, data mining techniques on all the gathered information can be used to improve predictive maintenance strategies, or to detect outliers, e.g., in energy usage in particular cars, or about certain driver styles. This centrally gathered and processed information, can thus be used to improve the overall fuel and maintenance efficiency of the fleet. 

Note that all of the above really changes the character of the cars, the car manufacturer and the transport company. Yes, it is still about steel, oil, rubber and cargo, but largely also about embedded ICT, data mining and information processing. From the perspective of the car manufacturer, the focus shifts from selling cars to services to sold cars. This is not new to the high-tech industry: there are more companies that focus more on the services they provide on top of their products, than on the actual sales of the products; the mobile telephony sector, but also the printing and copying sector are examples. And even a company like Rolls Royce does so for some of its aircraft engines; already in the 1960's they introduced their ``power-by-the-hour'' concept \cite{rolls-royce}.

\subsection{Situational awareness}
A final example considers so-called situational awareness, as worked upon in the ESI project Poseidon \cite{poseidon2013}. In this project, executed in cooperation with Thales, the aim was to develop an integrated coastal guard system, in which multiple sources of information are combined to provide the national coast guard with an integrated view on the coastal safety. This integrated view can be of use for (ship) collision avoidance purposes, for search-and-rescue operations, for maritime pollution response and for long-range detection and tracking of possibly unknown objects. The information to be integrated stems from many sources, both public and private, such as real-time measurements (radio, radar, sonar, satellite), public or private databases of vessel whereabouts (so-called AIS databases; see web-sites such as \url{www.fleetmon.com} or \url{www.vesselfinder.com}) or vessel history.  The data available will not always be the same: not all sources do provide information at all times, or about all vessels, the data formats are not a priori known and might change over time, and the data itself might be of varying quality and trustworthiness. Also, there might be attempts to seriously tamper the system, in that some parties might deliberately inject incorrect information into the databases or sent incorrect information via communication channels.  Overall, the goals is to built a robust and adaptable system, without a priori known configuration, that can be used to provide means for outlier detection. Clearly, each of the individual systems that is integrated does provide an independent functionality, but in total, in cooperation, a better and more reliable functionality can be attained.

\section{Implications for model-driven design}
\label{challenges}

What should become clear from the above considerations and examples is that future systems-of-systems will be very large, heterogeneous, will have partly unknown subsystems (from the viewpoint of other subsystems), will possibly vary their structure and cooperations over time, will certainly contain complex data dependencies, and the subsystems will have to be able to communicate with a changing set of partners. Still, systems-of-systems will in many cases underlie stringent requirements regarding performance and dependability. 

We advocate a model-driven approach to support the design of systems-of-systems. However, to deal with true systems-of-systems design, such a model-based approach has to be able to deal with largeness (scalability), heterogeneity (in model class), under-specification (black-box behaviour), time-varying models (in terms of model structure and model parameters), data dependencies and a large variability in data inputs. On top of that, in such an approach, due to the dynamic behaviour of systems-of-systems, notions of ``online design and integration'' (for new components coming in or being exchanged with older ones, while operation continues) have to be adequately modelled. This is a very challenging set of requirements, especially when seen in light of current-day modelling and analysis tools and techniques, that typically do not scale well, that require model-homogeneity and time-invariance, no data dependencies and static structures, to name a few.

We do not provide a new recipe, a new model type or class here, but instead list (non-exhaustively) a number of important conditions that need to be fulfilled in order for a model-based design method to be of true value in a systems-of-systems design context:
\begin{itemize}
\item We do not expect ``a single model class'' to be possible, nor to be useful. Much more, like systems-of-systems themselves, we believe that {\em cooperating models}, through well-defined interfaces, are the best way to support design processes. A nice example at stake has been put forward in the European Destecs project (see \url{www.detecs.org}) in which discrete-event models are combined with continuous (control) models. Approaches along these lines also cater for model inhomogeneity, thus allowing different design themes (disciplines) to employ their methods of choice.  
\item The importance of data dependencies in systems-of-systems make that fully analytical model solutions are beyond expectation. Instead, approaches that allow for {\em simulation} (or some hybrid form of simulation and other techniques) appear most fruitful.  Moreover, the rationalist's idea that one can design a complex system through pure thought (proof) alone, as is widespread in computer science, does not apply here\footnote{Did it ever apply?}. Instead, an approach stronger based on empiricism appears much more appropriate.
\item {\em Uncertainty} in the models, for instance about other parameters, other model components, etc., can be dealt with in several ways, such as probabilistically, stochastically, or using non-deterministic models. Very good progress has been made with probabilistic and stochastic models in recent years. Note that non-determinism is not compatible with simulation techniques. Uncertainty and sensitivity analysis can be employed to investigate the impact of parameter uncertainties.
\item To deal with situations in which complete subsystems have unknown structure and behaviour, approaches based on {\em model-mining} and {\em test-based modelling} appear useful to come up with overall behavioural models.
\item {\em Compositional modelling and analysis} are very strong techniques, however, these need to be enhanced towards extra-functional system characteristics, like performance and dependability. The ``good-old'' flow-equivalent server centre analysis \cite{CHW1975} developed in the realm of computer performance analysis in the 1970's serves as a good example.
\item For all modelling and analysis techniques to be developed, it has to be made sure that these can be used through state-of-the-art design tools as they are being used in {\em industrial practice}. It is naive to think that (industrial) system design engineers will acquire and adapt to academically developed tools and techniques, unless they are embedded in the (typically) company-prescribed design flow. 
\end{itemize}

\noindent Finally, the field of systems-of-systems design appears to be an excellent opportunity for computer scientists to team up with true system designers and system design approaches from, e.g., the aeronautics or automotive field. We should not shy away from these as being imprecise or too much engineering style; these methods have put men on the moon!
Knowledge of classical studies on design from these fields, like \cite{Brooks2010,Hitchins2007,maier2002,Simon1996,Vincenti1990}, might actually help to unleash the great potential of the powerful techniques and tools that have been developed over the last decades.

\nocite{henzinger-sifakis2007,ESI-SRA2012,BCG2004}
\bibliographystyle{eptcs}
\bibliography{aisos-bib}
\end{document}